\documentclass[letter,12pt]{article}
\usepackage[dvips]{graphicx}

\begin{document}
\baselineskip=14pt
\parindent=.2in

\begin{center}

{\large \bf Quantum Field Theories on a 
Noncommutative\\ 
Euclidean Space: Overview of New Physics}

\vspace{.4in}

Yong-Shi Wu\\

\vspace{.2in}

Department of Physics, University of Utah\\
Salt Lake City, UT 84112, U.S.A.

\vspace{.3in}

ABSTRACT\\

\vspace{.1in}

\begin{minipage}{4.5in}{\footnotesize
In this talk I briefly review recent developments 
in quantum field theories on a noncommutative 
Euclidean space, with Heisenberg-like commutation 
relations between coordinates. I will be 
concentrated on new physics learned from this 
simplest class of non-local field theories, 
which has applications to both string theory 
and condensed matter systems, and possibly 
to particle phenomenology.}
\end{minipage}

\end{center}

\vspace{0.15in}

{\large 1. Noncommutative Field Theories}
\vspace{.2in}

In this talk I will give a brief overview of 
the new physics recently learned from quantum 
field theories on a {\it noncommutative 
Euclidean} space. Below I will call them simply 
as {\it noncommutative} field theories (NCFT).
(Because of the time limitation, it is regretful 
that many significant topics and contributions 
in this field have to be left out.)

By definition, a noncommutative Euclidean 
space is a space with noncommuting spatial 
coordinates, satisfying the Heisenberg-like 
commutation relations:

$$
[x^i, x^j]=i\theta^{ij}, 
\eqno(1)
$$

\noindent where $\theta^{ij}=-\theta^{ji}$ 
are real constants. (There are other types of 
noncommutative spaces, with coordinates satisfying
Lie-algebraic or Yang-Baxter relations, or
with space-time noncommutativity. In this talk I 
am restricted only to the case of Eq. (1).) There 
are two ways of interpreting these commutation 
relations. 

The first way is to interpret $x^i$ as operators
in a Hilbert space satisfying Eq. (1). In this 
interpretation, the noncommutative space can be 
mathematically viewed as generalization of phase
space in usual quantum mechanics. The second 
is to interpret coordinates $x^i$ as functions 
that generate a noncommutative algebra of functions 
(or fields) on the space. One may develop a 
classical field theory from this point of view, 
by realizing the algebra of fields in the set of 
ordinary functions of commuting variables 
$\{x^\mu\}$ with the Moyal star product [1] 
(i.e. deformation quantization)  

$$
(f\star g) (x)= \exp \{\frac{i}{2}
\theta^{\mu\nu} \frac{\partial}
{\partial x^\mu} \frac{\partial}
{\partial y^\nu}\} f(x)g(y)|_{x=y}.
\eqno(2)
$$ 

\noindent (Here $\mu=0$ labels the time 
component, and we set $\theta^{0i}=0$.
It is easy to check that $x^\mu\star x^\nu - 
x^\mu\star x^\nu=i\theta^{\mu\nu}$.) Then we 
can transplant the usual differentiation and 
integration to the algebra of fields. So the 
(classical) action principle and the 
Euler-Lagrange equations of motion make sense. 
The key difference from the usual field theory 
is that everywhere the usual multiplication 
of two fields is replaced by the non-local
star product (2). For example, for a 
scalar $\phi^4$ theory, we have 

$$
{\cal S}= \int {1 \over 2} \partial_{\mu}
\phi \star \partial^{\mu} \phi + {1\over 2} 
m^2 \phi \star \phi + {g\over 4!} 
\phi\star\phi\star\phi\star\phi.
\eqno(3)
$$

\noindent For a gauge theory, the 
noncommutative Yang-Mills (NCYM) action is

$$
{\cal S}_{YM}= - \int {1 \over 4} 
F_{\mu\nu}\star F^{\mu\nu},\qquad
F_{\mu\nu}=\partial_{\mu} A_{\nu} 
-\partial_{\nu} A_{\mu}+ A_{\mu}
\star A_{\nu} - A_{\nu}\star A_{\mu}.
\eqno(4)
$$

\noindent Note that even $U(1)$ theory becomes
non-abelian, since $F_{\mu\nu}$ always contain
the star-commutator term. In the following I
will give a summary of new physic in this class 
of NCFT, with coordinate noncommutativity (1).  
Possible applications to the real world will 
be addressed in the last section.

\vspace{0.4in}
{\large 2. New Interaction Vertices}
\vspace{.2in} 

In the deformation quantization approach,
noncommutative quantum field theory (NCQFT) is
developed using path integral formalism. The 
propagator of a field is the same as usual, 
since ignoring boundary terms we have

$$ 
\int \phi \star \psi =\int \phi \psi.
\eqno(5)
$$ 

\noindent So changing the ordinary product 
to the star product only affects the 
interaction terms, giving rise to new 
vertices in NCQFT. 

In the NCFT (3), say, the $\phi^4$ 
interaction term contains an infinite 
number of higher-order derivatives. So it 
gives rise to {\it non-local} interactions 
characterized by a new momentum scale 
$\Lambda_{NC} = 1/\sqrt{\theta_{nc}}$. 
(Here we assume all $\theta^{ij}$ are 
of the same magnitude, denoted as 
$\theta_{nc}$.)  In path integral, 
the star product in this term generates 
a {\it momentum dependent phase factor} 
in the four-$\phi$ vertex:

$${g\over 6} \exp\{{i\over 2} 
\sum_{i<j} k_i \wedge k_j \},
\eqno(6)
$$ 

\noindent where $k_i \wedge k_j = \theta^{ij} 
k_i\, k_j$. Note that this vertex is 
invariant only under {\it cyclic} 
permutation of particle legs. This leads to 
the distinction between {\it planar} and 
{\it non-planar} diagrams in perturbation.

New interaction vertices appear in NCYM too.
According to the action (4), noncommutative 
quantum electrodynamics (NCQED) contains 
$3-photon$ and $4-photon$ vertices, both 
having a momentum dependent phase factor 
arising from the star product. Moreover, 
in NCQCD the star commutators in Eq. (4) 
does not close for $SU(3)$ group, so the 
latter has to be extended to $U(3)$. Also 
the $U(1)$ and $SU(3)$ subgroups do {\it 
not decouple}, and there are new type of 
vertices like $U(1)-SU(3)-SU(3)$ couplings 
etc. 

The effects of these new interactions are 
the origin of new physics in noncommutative
space, and they may be experimentally explored 
to probe possible coordinate noncommutativity 
in the real world.  
 
\vspace{0.4in}
{\large 3. UV-IR Entanglement}
\vspace{.2in} 

In perturbation theory, the non-locality gives 
rise to a novel entanglement of the UV and IR 
behavior. One manifestation of this effect is 
the one-loop non-planar correction to the 1PI 
two-point function [2]:

$$
\Gamma_{np}^{(2)}={g^2\over 96} [\Lambda_{eff}^2
-m^2\ln {\Lambda_{eff}^2 \over m^2} + O(1)],
\eqno(7)
$$

\noindent with $\Lambda_{eff}$ the effective 
cut-off which is related to the true cut-off 
$\Lambda$ by

$$\Lambda_{eff}^2={1 \over \Lambda^{-2}+
p\circ p},
\qquad p\circ q=- p_i (\theta^2)^{ij} q_j.
$$

\noindent It is obvious that the UV limit 
$\Lambda\to\infty$ and the IR limit $p\to 0$ 
do {\it not commute}, so that {\it the UV and 
IR behavior are entangled!} 

The UV-IR entanglement is a novel, essential 
and far-reaching feature of any NCFT. 
This is because of the non-local nature of 
interactions associated with coordinate 
noncommutativity: One has the space-space
uncertainty relations like

$$
\Delta x^1 \Delta x^2
\sim \theta^{12}.
\eqno(8)
$$ 

\noindent Assume $\theta^{12}\ne 0$. If one 
squeezes $\Delta x^1 \to 0$, then $\Delta x^2$ 
must tend to $\infty$ and {\it vice versa}. 

Some researchers feel very uncomfortable with the
UV-IR entanglement and are worried that it may 
make the renormalization group (RG) not work. 
Indeed some features of RG in {\it local} 
field theories should not work in NCFT, which 
are known to be non-local. But this does {\it 
not} imply that RG should not work (see Sec. 8
below). Of course, how RG survives in NCFT 
should be studied carefully case by case.

\vspace{1.0in}
{\large 4. Noncommutative Solitons}
\vspace{.2in}

Another piece of significant new physics in
NCFT is the existence of {\it new static 
solitons}. For example, in ordinary scalar 
theories, there is a famous Derrick's 
theorem: Static solitons exist only in space 
with dimension $d=1$. This can be understood 
by a simple scaling argument: If one shrinks 
the size of the soliton by changing ${\bf x} 
\to \lambda^{-1}{\bf x}$, then the kinetic 
energy scales as $K \to \lambda^{2-d} K$, 
while the scalar potential scales like 
$V\to \lambda^{-d} V$. In $d\ge 2$, a soliton 
configuration can reduce its energy by 
shrinking to zero size. 

Obviously such an argument is {\it inapplicable} 
to noncommutative space, in which a point does not 
make sense and there is a natural {\it minimum 
length scale}, given by $\sqrt {\theta_{nc}}$.
Mathematically this is because $V(\phi)$, as a
star polynomial, contains derivatives of $\phi$, 
so shrinking the size may lead to an increase, 
rather than decrease, in interaction energy. The 
existence of static scalar solitons in $d\ge 2$ 
is shown explicitly in ref. [3]. For example, 
in the large $\theta$-limit, the scalar potential 
dominates and a static soliton can be generated 
from a {\it projector function} satisfying

$$
\phi_0(x) \star \phi_0 (x) = \phi_0 (x),
\eqno(9)
$$

\noindent which can be easily solved in $d=2$.
The correspondence between noncommutative fields
and operators in Hilbert space is extremely 
useful in generating multi-soliton solutions. 
Physically the new scalar solitons represent 
bubbles of false vacuum with size set by 
$\sqrt{\theta_{nc}}$. They are stable when the 
potential $V(\phi)$ has at least two minima. 

New solitons also appear in noncommutative 
gauge theories. In particular, there are 
{\it nonsingular} $U(1)$ monopoles [4] and
$U(1)$ instantons [3,5]. Moreover, the 
small-instanton singularity in the moduli 
space of ordinary instantons gets resolved in 
NCYM, since there exists a minimum size for 
instantons, set by $\sqrt{\theta_{nc}}$ [5].

The existence of new solitons dramatically 
changes the spectrum, and greatly enriches 
nonperturbative physics of the theory.  

\vspace{0.4in}
{\large 5. Spacetime Symmetry Breaking}
\vspace{.2in} 

The third significant new feature of an NCFT
is the {\it natural breaking of spacetime 
symmetry} by coordinate noncommutativity. 
This is because the noncommutativity parameters 
$\theta^{ij}$ behave like {\it a background in 
space.} In $3+1$ dimensional NCQED, $\theta^{ij}$
is unchanged under ${\bf x}\to - {\bf x}$, 
parity $P$ is invariant, while $C$ and $T$ are
non-invariant because they lead to $\theta 
\to -\theta$. However, $CPT$ is still a 
symmetry[6], $CP$ is broken too. Physically 
these could be understood in the following 
way: an electron in NCQED has a tree-level 
momentum dependent electric dipole moment, 
given by
  
$${\bf \mu_e}=-{e\over 4\hbar} 
({\bf \theta} \times {\bf p}). 
\eqno(10)
$$ 

\noindent It is easy to verify the above 
statements on discrete symmetries for this 
expression.  

Moreover, the $\theta$-background makes {\it 
space anisotropic} and {\it breaks Lorentz boost} 
symmetry in the active sense, namely if one rotates 
or Lorentz-boosts a physical system, its behavior
will become different. Of course, if one makes 
passive rotational and boost transformations 
of coordinates, the physics would be unchanged. 
Therefore, the effects of space-time symmetry
breaking effects can be used to probe the 
coordinate noncommutativity parameters 
$\theta^{ij}$ or, at least, to set observational 
upper limits for them. (See Sec. 7 below.)
 
\vspace{0.4in}
{\large 6. New Physics from Radiative Corrections}
\vspace{.2in}

Since the amplitude of {\it planar} diagrams in 
NCQFT differs from the ordinary cases by an overall
phase factor that depends merely on external momenta, 
introducing coordinate noncommutativity does 
not make a $UV$ divergent ordinary field theory
finite. However, in NCFT non-planar diagrams  
are suppressed due to loop-momentum dependent 
phase factors, and there are new interaction 
vertices due to star product, so the $UV$ 
behavior and radiative corrections are affected. 

The renormalizablility, i.e. the counterterms being
of the same form and with the same $\theta$ parameter(s) 
in the star product as in the bare action, has been 
verified explicitly at least at one loop for many 
NCFT's. The most dramatic change of the $UV$ 
behavior occurs in $U(1)$ NCYM theory in $3+1$ 
dimensions, which is {\it asymptotically free}. 
The one-loop $\beta$-function of $U(N)$ NCYM 
is given by [7]

$$
\beta (g) = - {11\over 3} {g^3 N\over 
(4\pi)^2},
\eqno(11) 
$$ 

\noindent which is valid even for $N=1$! 

Similarly $U(1)$ noncommutative Chern-Simons 
(NCCS) theory in $2+1$ dimensions has a 
non-vanishing one-loop level shift, which 
shows up in ordinary Chern-Simons theory 
only for $SU(N)$ with $N\ge 2$. The one-loop
level shift in $U(N)$ NCCS is given by [8]

$$
k\to k+ N {\it sign} (k),
\eqno(12)
$$

\noindent which is valid also for $N=1$!
In ordinary $U(1)$ Chern-Simons theory,
even with coupling to matter, there is 
no such shift at least at one loop [9].
The result (12) indicates that the {\it 
topology of the gauge group} on noncommutative 
space should be different from that in 
ordinary space.

\vspace{1.0in}
{\large 7. Observational Limits}
\vspace{.2in}

Whether our real world is a noncommutative space
or not can be determined only by observations. 
We need to probe new physics brought up by NCQFT.
Here we cite two methods as examples; the second
one gives the best observational upper limit for 
the noncommutativity parameter(s) $\theta_{nc}$. 

One method is to explore the Lamb shift of the 
hydrogen atom. In NCQED, there is a tree-level
electric dipole moment for the electron. Both the 
electric and magnetic dipole moment receive 
one-loop radiative corrections. Their contributions 
to the $2S_{1/2}-2P_{3/2}$ hyperfine splitting are
estimated in ref. [10]. Comparing with the experimental
error bar for the most recent precise measurement
gives the following observational limit:

$$
|\theta_{nc}|\le (0.1\, TeV)^{-2}.
\eqno(13)
$$

Another method is to probe possible effects on 
space anisotropy and Lorentz violation due to 
$\theta^{ij},\;(i,j=x,y,z)$, which look like 
a background. More concretely, there should 
be change in clock rate due to rotation of 
the Earth. An atomic clock comparison experiment 
was carried out, quite a bit time ago [11], to 
monitor closely the difference between two atomic 
clocks and to search for variations as the Earth 
rotates. New analysis of the old data has been 
done from the point of view of NCQED [12], which 
sets the upper limit

$$
|\theta^{YZ}, \theta^{ZX}| \le (10\, TeV)^{-2}.
\eqno(14)
$$

\noindent Here $X,Y,Z$ are refereed to the 
non-rotating celestial equatorial coordinates.

\vspace{0.4in}
{\large 8. Renormalization Group and Critical 
Exponents}
\vspace{.2in}

In condensed matter physics, renormalization 
group equations (RGE) provide a systematic and 
powerful tool to study the low energy or large 
distance behavior of a many-body system, in 
particular the critical behavior near a second 
order (or continuous) phase transition point. 
A prototype of such phase transition occurs 
in the Landau-Ginzburg model, namely a real 
$\phi^4$ theory with a ``wrong-sign'' mass term, 
which leads to spontaneous symmetry breaking. In
ordinary space, depending on the sign of $m^2$ 
there are only two phases possible: With $m^2>0$,
the system is in a {\it disordered} phase with 
$<\phi>=0$, while for $m^2<0$ it is in an {\it 
ordered} phase with $<\phi>\ne 0$, with a {\it 
phase transition} at $m^2=0$. For the critical 
behavior near the phase transition, it is
well-known that $D=4$ is the critical dimension: 
Above it (for $D>4$), mean field theory is valid,
and below it ($D<4$) one needs to go beyond mean 
field theory and exploit the RGE to get correct 
critical exponents. 

With my postdoc, G. H. Chen, we have studied 
[13] the phase diagram and phase transitions 
in the noncommutative Landau-Ginzburg model 
(NCLGM), described by the action (3). This 
model is known to be one-loop renormalizable, 
and the counterterm of the $\phi^4$ 
interaction has a star product with the
same $\theta$-parameters. Therefore, 
the $\theta$-parameters, that characterize
noncommutative geometry of the space, are 
{\it not renormalized}. Though this is
natural from the geometric point of view, 
it is contrary to the intuition from local 
quantum field theory, which considers 
coordinate noncommutativity as short-distance 
effects, which should be washed out at large 
distances. In our opinion, this intuition 
is {\it not} justified in noncommutative 
space, because of the {\it non-local $UV-IR$ 
entanglement}: The large-distance behavior 
of the system does carry fingerprints of the 
noncommutative geometry at short distances.

We exploited the so-called {\it functional} RGE 
approach [14] popular in condensed matter theory,
to overcome the difficulty due to the $UV-IR$
entanglement. The idea is that to derive RGE,
we only need to perform the integration over a 
{\it thin shell} in momentum space at one loop, 
which gives us the effects of changing the cutoff
from $\Lambda$ to $\Lambda-d\Lambda$. This 
has the advantage of avoiding possible IR 
singularity, best suited to our purpose. Using
this approach we obtained the RGE for the 
dimensionless variables $r\equiv m^2/\Lambda^2$ 
and the coupling constant $u\equiv g\Lambda^{4-D}$ 
as follows:

$$
{dr\over dt}=2r+{u\over 2}K_D (1-r),\qquad
{du\over dy}=(4-D)u-{3\over 2}K_D u^2,
\eqno(15)
$$

\noindent which are the same as in ordinary 
Landau-Ginzburg model (LGM), and there is no 
need to consider the RGE for $\theta_{nc}$.
Here $t=s-1$ is the RG flow parameter defined 
by $\Lambda\to \Lambda/s$. What is novel in the
noncommutaive case is there is a non-vanishing 
wave function renormalization from the non-planar 
part of the one-loop tadpole diagram for the 
propagator, which leads to a {\it negative} 
$\theta$-dependent anomalous dimension for 
the order parameter field $\phi$:

$$
\gamma= - {u\over 48} K_D \tilde{\theta}^2,
\eqno(16)
$$ 

\noindent with $\tilde{\theta}\equiv \theta 
\Lambda^2$ being dimensionless, and $K_D$ 
the area of unit sphere in $D$ dimensions.

For {\it small} values of $\tilde{\theta}_{nc}$ 
in dimension $D=4-\epsilon$, we still have the 
same Wilson-Fisher fixed point as before:
$r^*=-{\epsilon}/6, u^*=16\pi^2\epsilon/3$. 
However, the critical behavior near phase 
transition is altered in one aspect: The 
critical exponent $\eta$ becomes negative 
and $\tilde{\theta}$-dependent: 
$\eta=-\epsilon\tilde{\theta}^2/72.$ 
(In ordinary LGM, $\eta$ vanishes.)   

\vspace{0.4in}
{\large 9. Non-uniform Ordered Phase and the Lifshitz Point}
\vspace{.2in}

The negative anomalous dimension (16) leads to an 
instability for sufficiently {\it large} 
$\tilde{\theta}$ parameters, when it makes the 
total dimension of $\phi$ zero or negative. In other 
words, when $\tilde{\theta}\ge\tilde{\theta}_{c}$,
e.g. at the WF fixed point $\tilde{\theta}_{c} 
=12/\sqrt{\epsilon}$, the sign of the 
quadratic kinetic term will become negative, 
signaling a {\it momentum-space instability}. 

In such a case, one has to include a positive 
{\it fourth-order} derivative term to maintain 
the stability of the system. (We have verified 
that if at tree level there is no such term, 
it will be induced at one loop.) Then the 
low-$r$ (i.e. low temperature) ordered phase 
becomes {\it non-uniform} because of the 
instability: The order parameter now must 
be modulating in space:

$$ \phi({\bf x})=\phi_0 \cos ({\bf k}\cdot {\bf x}),
\eqno(17)
$$

\noindent with a wave vector $k$ at the 
minimum of the combined kinetic energy.
 
Therefore, the phase diagram of the NCLGM in 
$r-\tilde{\theta}$ space is complicated, as shown in 
Fig. 1. There are {\it three} possible phases:
For positive $r$, the system is always in 
disordered phase. However, for large and 
negative $r$, the ordered phase can be 
{\it uniform} or {\it non-uniform}, a sort of 
{\it striped phase}, depending on the value 
of $\tilde{\theta}$.

\begin{figure}
\begin{center}
\includegraphics[width=3in,height=6in,keepaspectratio]{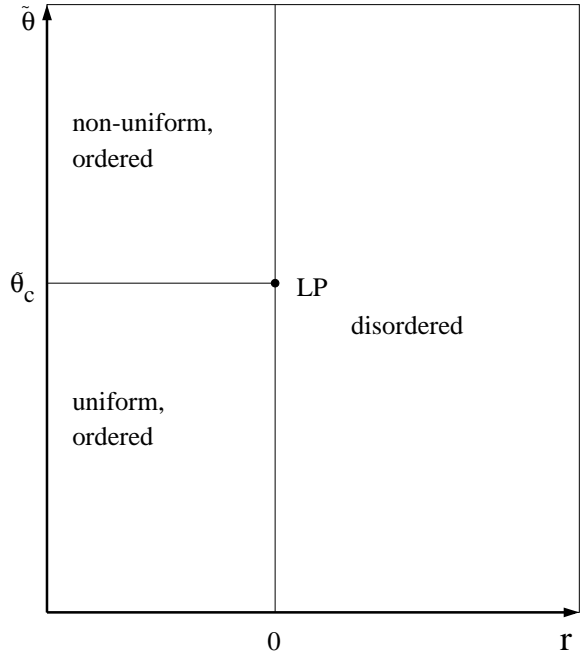}
\caption{The Lifshitz point in the phase diagram of NCLGM}
\label{fig:phase_diagram}
\end{center}
\end{figure}

On the ordered phase side, the transition from 
a uniform to a non-uniform phase is of first order. 
On the other hand, the transition from a disordered 
to an ordered phase (by tuning $r$) is always of 
second order. Therefore there is a {\it Lifshitz} 
critical point [15] at the intersection of the 
first and second phase transition lines. (A similar 
scenario has been proposed earlier with a different 
argument in ref. [16].) We have used a mean field 
theory to determine the critical behavior near the 
Lifshitz point. For details, we refer the interested 
reader to the original paper [13]. Our feeling is 
that the appearance of a {\it non-uniform, ordered} 
phase at large values of $\tilde{\theta}$ should be 
a general feature of noncommutative space. 

\vspace{0.4in}
{\large 10. Applications or Realizations}
\vspace{.2in}

To conclude my talk I discuss applications or 
realizations of NCFT in physics. 

One realization is known to be the lowest 
Landau level (LLL) in the {\it quantum Hall} 
systems: In some semi-conductor devices, 
electrons may be constrained to moving in 
a plane. In a strong transverse magnetic 
field and at very low temperature, 
electrons may be further restricted to 
the LLL, in which the cyclotron motion 
has a fixed minimal radius. It is 
well-known that the guiding center 
coordinates of the cyclotron orbit in 
the LLL do {\it not} commute [17]:

$$
[X, Y] =i {\hbar c \over e B}.
\eqno(18)
$$

\noindent This is obviously a realization 
of Eq. (1). So it is generally believed  
that NCFT should be useful to the theory of
quantum Hall systems. However, nobody has 
been able to {\it discover new physics} 
directly with the help of NCFT.   

More applications can be found in {\it string 
theory}. For example, the coordinates of an 
open string endpoint on a D-brane, in an 
anti-symmetric tensor $B$-background, are 
shown to be noncommuting [18], with a form 
similar to Eq. (18). If our observed world 
happens to be on such a D3-brane, then the 
NCQFT on space (1) would become relevant 
to the real world, particularly to {\it 
particle physics phenomenology}! Another 
application is to Witten's second quantized 
{\it string field theory} [19], in which 
the product between two string fields is 
noncommutative. Recently it has been shown 
[20] that the string-field product can be 
rewritten as an infinite tensor product of 
Moyal's star products in single string 
Hilbert space. Thus, the techniques and 
physics of NCFT are expected to be very 
useful in string field theory. 

For physics readers interested in NCFT,
there is an excellent review paper in 
Review of Modern Physics [21]. Also a
reprint volume of original papers in 
NCFT has been published by the Rinton 
Press [22]. Most papers quoted
here can be found in that volume.

\vspace{0.4in}
{\large Acknowledgements}
\vspace{.2in}

The author thanks the organizers of TH-2002, Prof.
J. Zinn-Justin and Prof. D. Iagolitzer for inviting 
him to speak at this wonderful conference. He also
acknowledges pleasant collaboration with Dr. 
Guang-Hong Chen. This work was supported in part 
by U.S. NSF grant PHY-9907701.

\vspace{1.0in}

{\large References}
\vspace{.2in}

\noindent 1.  J.E. Moyal, {\it Proc. Cam. Phil.
Soc.} {\bf 49}, 45 (1949).  \\
2. S. Minwalla, M. V. Raamsdonk and N. Seiberg, 
{\it JHEP} {\bf 0002}, 020 (2000).\\
3.  R. Gopakumar, S. Minwalla and A. Strominger,
{\it JHEP} {\bf 0005}, 020 (2000).\\
4.  N. Nekrasov and A. Schwarz, {\it Commu.
Math. Phys.} {\bf 198}, 689 (1998). \\ 
5.   D. Gross and N. Nekrasov, {\it JHEP} {\bf 0007},
  034 (2000). \\ 
6.   M.M. Sheikh-Jabbari {\it Phys. Rev. Lett.} {\bf 84}, 
  5265 (2000).\\
7.   A. Armoni, {\it Nucl. Phys.} {\bf B593}, 229 (2001).\\
8.   G.H. Chen and Y.S. Wu, {\it Nucl. Phys.} {\bf B593},
  562 (2001). \\
9. See, e.g., G.W. Semenoff, P. Sodano and Y.S. Wu,
{\it Phys. Rev. Lett.} {\bf 62}, 715 (1989), and 
references therein. \\
10.   M. Chaichian, M.M. Sheikh-Jabbari, and A. Tureanu,
{\it Phys. Rev. Lett.} {\bf 86}, 2716 (2001). \\ 
11.   J.D. Prestage {\it et al.}, {\it Phys. Rev. Lett.}
{\bf 54}, 2387 (1985). \\
12.   S.M. Carroll, J.A. Harvey, V.A. Kostelecky, C.D. Lane, and 
T. Okamoto, {\it Phys. Rev. Lett.} {\bf 87}, 141601 (2001).\\
13.   G.H. Chen and Y.S. Wu, {\it Nucl. Phys.} {\bf B622},
 189 (2002). \\
14.   R. Shanker, {\it Rev. Mod. Phys.} {\bf 66},
  129 (1994). For a particle physicist's formulation, 
  see J. Polchinski, "{\it Effective Field Theory 
  and the Fermi Surface}", preprint UTTG-20-92 and 
  in Proceedings of TASI 92. \\
15.   See, e.g., I.D. Lawrie and S. Sarbach, in {\it Phase 
Transition and Critical Phenomena}, vol. 9. (Academic 
Press; 1984).\\  
16.   S. Gubser and S.Sondhi, hep-th/0006119.\\
17.   See, e.g., F.D.M. Haldane and Y.S. Wu, {\it Phys. Rev.
Lett.} {\bf 55}, 2887 (1985).\\
18.   See, e.g., N. Seiberg and E. Witten, {\it JHEP}
{\bf 9909}, 032 (1999) and references therein.\\
19.   E. Witten, {\it Nucl. Phys.} {\bf B268}, 253 (1986).\\
20.   I. Bars, {\it Phys. Lett.} {\bf B517}, 436 (2001);
   M.R. Douglas {\it et al.}, {\it JHEP} {\bf 0204}, 
022 (2002);
   Y.S. Wu and T.L. Zhuang, hep-th/0211199.\\
21.   M. Douglas and N. Nekrasov,  {\it Rev. Mod. Phys.} 
{\bf 73}, 977 (2001).\\ 
22.   "{\it Physics in a Noncommutative World, vol I. Field Theory}",
Ed. Miao Li and Y.S. Wu (Rinton Press; New Jersey; 2002).\\

\end{document}